\documentclass[aip,jcp,amsmath,amssymb,reprint,numeric]{revtex4-1}
\usepackage{graphicx}
\usepackage{subfigure}
\usepackage{color}
\usepackage{amsmath}
\usepackage{bm}
\usepackage{multirow}

\usepackage{ulem} %to cancel lines

\begin{document}

\title {Analysis of the anomalous mean-field like properties of Gaussian core model in terms of entropy}
%\author{Manoj Kumar Nandi} 
%\affiliation{\textit{{Polymer Science and Engineering Division, CSIR-National Chemical Laboratory, Pune-411008, India}}}
%\author{Sarika Maitra Bhattacharyya}
%\email{mb.sarika@ncl.res.in}
%\affiliation{\textit{{Polymer Science and Engineering Division, CSIR-National Chemical Laboratory, Pune-411008, India}}}
\author{Manoj Kumar Nandi} 
%\email{mb.sarika@ncl.res.in}
\affiliation{\textit{{$^{1)}$ Polymer Science and Engineering Division, CSIR-National Chemical Laboratory, Pune-411008, India}}}
%\email{mb.sarika@ncl.res.in}
\author{Sarika Maitra Bhattacharyya}
\email{mb.sarika@ncl.res.in}
\affiliation{\textit{{$^{1)}$ Polymer Science and Engineering Division, CSIR-National Chemical Laboratory, Pune-411008, India}}}
%\affiliation{\textit{{Polymer Science and Engineering Division, CSIR-National Chemical Laboratory, Pune-411008, India}}}

\date{\today}

\begin{abstract}
Studies of the Gaussian core model (GCM) have shown that it behaves like a mean-field model and the properties are quite different 
from standard glass former. In this work, we investigate the entropies, namely the excess entropy ($S_{ex}$)
and the configurational entropy ($S_c$) and their
different components to address these anomalies. Our study corroborates most of the earlier observations and also sheds new light
on the high and low
temperature dynamics. We find that unlike in standard glass former where high temperature dynamics is dominated by two-body 
correlation and low temperature by many-body correlations, in GCM both high and low temperature dynamics are dominated by many
body correlations. We also find that the many-body entropy which is usually positive at low temperatures and is associated with
activated dynamics is negative in GCM suggesting suppression of activation. Interestingly despite suppression of activation
the Adam-Gibbs (AG) relation which
describes activated dynamics holds in GCM, thus suggesting a non-activated contribution in AG relation. We also find an overlap 
between
the AG and mode coupling power law regime leading to a power law behaviour of $S_c$. 
From our analysis of this power law behaviour we predict
that in GCM the high temperature dynamics will disappear at dynamical transition temperature and below that, there
will be a transition to the activated regime. Our study further reveals that the activated regime in GCM is quite narrow.

\end{abstract}
\maketitle
\section{Introduction}

On cooling a liquid sufficiently fast it does not get enough time to crystallize and enters the supercooled liquid state.
On further cooling it becomes a glass.
The manner in which such supercooled liquid becomes amorphous rigid solid is poorly understood. Numerous theories have been
proposed to explain this slowing down of the dynamics in supercooled liquids \cite{usoft1,usoft2,usoft3,usoft4} but none of them have 
successfully answered all the questions.
 Mode coupling theory
(MCT), known as the microscopic theory of glass transition is one such theory \cite{meanfield_gcm_kuni6}. According to the predictions of
this theory, at the dynamical transition temperature, the relaxation time diverges in a power law manner
\cite{gotze1999,GMCT_Reichman}.  This power law behaviour is indeed observed in many experiments and computer simulation studies
\cite{du_mct_exp,RFOT,cavagna2001_epl,sciortino-pre-2002,sciortino-saddle}. However for these systems at low enough temperatures
one observes a departure from the power law and the divergence predicted is thus avoided.

According to the random first order transition theory (RFOT), at the dynamical transition temperature the system 
 is trapped in one of the basins of its rugged free energy landscape 
\cite{RFOT_cavagna}. %, RFOT_cavagna2}. 
At the 
mean-field level, the system is permanently trapped in one such minima as the barriers between 
the minima
become infinite and thus as predicted by MCT 
the dynamics is completely frozen. In finite
dimensions the barrier heights are less, thus this transition predicted by MCT is suppressed by the activation process. 
At low temperatures the dynamics is 
governed by activation and relaxation time follows the well-known Adam-Gibbs (AG) relation \cite{adam-gibbs},
which expresses relaxation time, $\tau$ in terms of a thermodynamic quantity, the configurational entropy $S_c$.

Although MCT like power law behaviour is found in simulation and experimental studies, thus predicting a transition temperature, 
$T_c$, the microscopic MCT when solved numerically
using structural information from simulations, predicts a transition at $T_c^{micro}$ which is higher than $T_c$
\cite{GMCT_Reichman,sciortino-pre-2002}. The reason for the prediction of
this higher transition temperature is not fully understood. However it is 
believed that the Gaussian approximation made in the naive form of MCT \cite{role_pair} which leads to
 the non-linear feedback mechanism is responsible for the higher value of $T_c^{micro}$.
Also in a recent work we have shown that the form of the vertex
function which depends on the structure factor might also be responsible for this premature divergence \cite{unravel}.

 As MCT is a microscopic mean-field theory, it is expected that the predictions made by MCT should systematically improve as we 
go towards mean-field like systems by increasing the dimension. It was found that for 4 dimensional hard sphere 
fluid, MCT predicts the slow dynamics much 
better than it does for lower dimensions \cite{4d_mct}. Another way of achieving mean-field
effect is by making the interaction between the particles long range. 
 Ikeda {\it {et al.}} have shown that
the Gaussian core model (GCM) behaves more like
a mean-field system \cite{ikeda2011slow,ikeda_Japan}.
The discrepancy between $T_c$ and $T_c^{micro}$ is around $20\%$
for GCM whereas for standard glass former like Kob-Andersen (KA) model it is above $100\%$. 
There are also other observations where GCM was found to behave quite differently from standard glass forming systems 
\cite{meanfieldGCM}.
It has been observed that for most of the glass former, the MCT power law exponent $\gamma$ varies 
when it is obtained from power law fits of relaxation time and diffusion coefficient but for GCM these values
come closer \cite{meanfieldGCM}. In standard glass former as the system approaches low temperatures, both the non-Gaussian parameter $\alpha_2(t)$
and the four-point correlation function $\chi_4(t)$ increase in a similar fashion. However in GCM the $\alpha_2(t)$ was found to 
grow much less than in KA model \cite{ikeda2011slow} but the $\chi_4(t)$ was found to grow much more \cite{meanfieldGCM}.
This apparent contradiction was explained in terms of mobility field. Large values of $\alpha_2(t)$ in KA model indicates  
large displacement of individual mobile particles whereas the enhancement of $\chi_4(t)$ in the GCM implies that this system has more cooperative
motion. From mode localization analysis it has been found that as temperature decreases the unstable directions that disappear at 
the dynamic transition temperature are highly delocalized for the GCM whereas they are increasingly
localized for other standard glass former like KA model \cite{meanfieldGCM_47}.
It has been shown that due to the high energy barriers, the hopping like motions
are strongly suppressed in GCM and the van Hove correlation function does not show any bimodal distribution even at
low temperatures \cite{meanfieldGCM}.

In this paper, we present a comparative study between KA binary mixture at density $\rho=1.2$ and mono-atomic
GCM at $\rho=1.5$ and 2.0. Our study is based on the calculation of entropy and its separation into different components and studying
its correlation with dynamics. We find that just like the other properties, the entropy and its components in GCM behave in a different
way when compared to KA model. In our study, we show that both high and low temperature dynamics in GCM is dominated by many-body
correlations and there is a suppression of activation. Surprisingly we find that even though there is a suppression of activation
the AG relation is valid. We also find that there is an overlap between the AG and MCT regime and from our analysis we can predict
that at a temperature, lower than that presented in this study the system makes a transition to activation dominated dynamics.

 The paper is organized as follows: The simulation details for various systems are given in Sec. II. In the next section, we describe
 the methods used for evaluating various quantities and provide other necessary backgrounds. In Sec. IV, we present the results and 
 discussions. Sec. V contains the conclusion.

\section{Simulation Details}

In this study, we perform extensive molecular dynamics simulations of two different glass forming liquid models.  
One is the binary Kob-Andersen Lennard-Jones liquid \cite{kob} and the other is a 
soft Gaussian core model \cite{ikeda2011slow}. The first system is binary and 
 the second is a monodisperse system. The total number 
density is fixed at $\rho=N/V$ with the total number of particles N (where $N = N_A + N_B$ for binary system) and system volume $V$.
 The molecular dynamics (MD) simulations are carried out using the LAMMPS 
package \cite{lammps}. 
 We perform the simulations in the canonical ensemble (NVT) using  Nos\'{e}-Hoover 
 thermostat.  %with integration time step 0.005 for KA model and 0.2 for GCM. %$\tau$.
 The time
constant for  Nos\'{e}-Hoover thermostat is taken to be 100  time steps.
The sample is kept in a cubic box with periodic boundary condition.
For all state points, three to five independent samples with run lengths $>$ 100$\tau$ ($\tau$ is the $\alpha$-
relaxation time) are analyzed.

\subsection{Binary mixture of Kob-Andersen Lennard-Jones particles}

The most well-known model for glass forming liquids is Kob-Andersen model which is a binary mixture (80:20) \cite{kob}.
 The interatomic pair  
potential between species $\alpha$ and $\beta$, with ${ \alpha,\beta}= A,B$, 
$U_{\alpha\beta}(r)$ is described by a shifted and truncated Lennard-Jones potential, as given by:
\begin{equation*}
 U_{\alpha\beta}(r)=
\begin{cases}
 U_{\alpha\beta}^{(LJ)}(r;\sigma_{\alpha\beta},\epsilon_{\alpha\beta})- U_{\alpha\beta}^{(LJ)}(r^{(c)}_{\alpha\beta};\sigma_{\alpha\beta},\epsilon_{\alpha\beta}),    & r\leq r^{(c)}_{\alpha\beta}\\
   0,                                                                                       & r> r^{(c)}_{\alpha\beta}
\end{cases}
\label{LJ_pot}
\end{equation*}

\noindent where $U_{\alpha\beta}^{(LJ)}(r;\sigma_{\alpha\beta},\epsilon_{\alpha\beta})=4\epsilon_{\alpha\beta}[({\sigma_{\alpha\beta}}/{r})^{12}-({\sigma_{\alpha\beta}}/{r})^{6}]$ and
 $r^{(c)}_{\alpha\beta}=2.5\sigma_{\alpha\beta}$. Length, temperature and
time are given in units of $\sigma_{AA}$, ${k_{B}T}/{\epsilon_{AA}}$ and $\surd({m_A\sigma_{AA}^2}/{\epsilon_{AA}})$, 
respectively.  The interaction parameters for Kob-Andersen model  
are,  $\sigma_{AA}$ = 1.0, $\sigma_{AB}$ =0.8 ,$\sigma_{BB}$ =0.88,  $\epsilon_{AA}$ =1, $\epsilon_{AB}$ =1.5,
 $\epsilon_{BB}$ =0.5, $m_{A}$ = $m_B$=1.0. 
% We perform MD simulations in the canonical ensemble (NVT) using  Nos\'{e}-Hoover 
% thermostat  with integration time step 0.005. %$\tau$.
% The time
%constant for  Nos\'{e}-Hoover thermostat is taken to be 100  time steps.
%The sample is kept in a cubic box with periodic boundary condition.
The integration time step is fixed at 0.005.
 System size is $N = 500$, where $N_A = 400$ and $N_B=100$ and the density of the system is $\rho=1.2$.

\subsection{Gaussian core model}

The  Gaussian core model  is a one-component system.  
The interaction potential is a Gaussian shaped repulsive potential. The potential is shifted and truncated at the cutoff 
$r^{(c)}=5\sigma$ and 
is given by:
\begin{equation}
 U(r)=
\begin{cases}
 \epsilon_0 \exp[-(r/\sigma)^2]-\epsilon_0 \exp[-(r^{(c)}/\sigma)^2],    & r\leq r^{(c)}\\
   0,                                 & r > r^{(c)}.
\end{cases}
\end{equation}
 Length, temperature and time are given in units of $\sigma$, ${k_{B}T}/{\epsilon_0}$ and $\surd({m\sigma^2}/{\epsilon_0})$, 
respectively.    
The interaction parameters are $\sigma$ = 1.0, $\epsilon_0$ =1.0, $m$ =1.0. 
The integration time step is fixed at 0.2.
 System size is $N = 3456$ and we choose two densities, $\rho=1.5$ and $\rho=2.0$ for our study.

\section{Definations}
\subsection{Relaxation time}% $\tau_{\alpha}$}
The relaxation times are obtained from the decay of the overlap function, $q(t)$, using the definition $q(t=\tau)=1/e$.
The overlap function $q(t)$ is defined as,
\begin{eqnarray}
 q(t) \approx \frac{1}{N}\left \langle \sum_{i=1}^{N} \Theta (\mid{\bf{r}}_i(t_0)-{\bf{r}}_i(t+t_0)\mid) \right \rangle \nonumber\\
\Theta(x) = 1, x \leq {\text{a implying “overlap”}} \nonumber\\
=0, \text{otherwise}.
\end{eqnarray}
The cut off parameter `a' is chosen as 0.3.

\subsection{Excess Entropy}
The thermodynamic excess entropy, $S_{ex}$, which arises
due to structural correlations is the difference between the total entropy $S_{total}$ and the ideal gas value $S_{id}$ at the same state
point (T,$\rho$). $S_{ex}$ is calculated by using the method described in Ref.\cite{sciortino2000thermodynamics}. The entropy is first 
evaluated at a state point, usually at a high temperature and low density, where the system behaves 
like an ideal gas. Relative to this state point, entropy at any other state point can be calculated by using the combination of 
an
isothermal and an isochoric path, avoiding any phase transition along this path. Along the isothermal 
path entropy change is given by,
\begin{equation}
 S(T,V')-S(T,V)=\frac{U(T,V')-U(T,V)}{T}+\int_V^{V'} \frac{P(V)}{T}dV,
\end{equation}

and along the isochoric path it is,
\begin{equation}
 S(T',V)-S(T,V)=\int_T^{T'}\frac{1}{T}\Big(\frac{\partial U}{\partial T}\Big)_V dT.
\end{equation}

\subsubsection{Pair and higher order excess entropy}
By using
Kirkwood factorization \cite{kirkwood} of the N-particle distribution function \cite{green_jcp,raveche,Wallace},
the excess entropy $S_{ex}$ can be expanded in an infinite series, 
\begin{equation}
S_{ex}=\sum_{n=2}^\infty S_n= S_2+\Delta S,
\label{deltas_sexc}
\end{equation}
 $S_n$ are partial entropies
which can be obtained by a suitable re-summation of spatial density correlations involving n-particle multiplets.
The pair excess 
entropy $S_2$ for binary system reads as,
\begin{equation}
 \frac{S_2}{k_B}=-2\pi\rho \sum_{\alpha \beta}x_{\alpha}x_{\beta}\int_0^\infty {g_{\alpha \beta}(r)\ln g_{\alpha \beta}(r)
 -[g_{\alpha \beta}(r)-1]}r^2dr,
 \label{s2}
\end{equation}
where $g_{\alpha \beta}(r)$ is the atom-atom pair correlation function between type $\alpha$ and type $\beta$, $\rho$ is the 
density of the system, $x_{\alpha}$ is the mole fraction of type $\alpha$ and $k_B$ is the Boltzmann constant.  
$\Delta S=S_{ex}-S_2$, is called the residual multiparticle entropy (RMPE) which contains the higher order contributions (beyond 
two-body) to 
the excess entropy \cite{rmpe_saija1}.

%============================================================
\subsection{Configurational Entropy}
The configurational entropy ($S_c$ ) per particle, is calculated \cite{srikanth_PRL}
by subtracting the vibrational entropy from the total entropy of the system :
$S_c (T ) = S_{id} (T )+ S_{ex}(T) - S_{vib} (T )$ \cite{shila-jcp,Srikanth_nature}. Here $S_{id}$ is the ideal gas entropy and the 
excess entropy, $S_{ex}$ is obtained by the method described in Sec. IIIB.
For vibrational entropy we use a harmonic approximation to the potential energy about a given local minima.
The detailed procedure for generating the local minima and calculating the vibrational entropy is given in Ref. 
\cite{srikanth_PRL,shila-jcp,Srikanth_nature}. 
As mentioned in an earlier study \cite{meanfieldGCM} in the calculation of the density of states, we also find some imaginary modes ($\sim 0.19\%$)
which we ignore to calculate the vibrational entropy. We believe that this will not make any change in the physical properties of
the system.

\subsubsection{Pair Configurational Entropy}
To obtain an estimate of the configurational entropy as predicted by the pair correlation we rewrite $S_{c}$ in terms of the 
pair contribution to 
configurational entropy, $S_{c2}$\cite{atreyee_prl},
\begin{equation}
S_{c}=S_{id}+S_{ex}-S_{vib}=S_{id}+S_{2}+\Delta S-S_{vib}=S_{c2}+\Delta S.
\label{sc2}
\end{equation}
\noindent
Where $S_{c2}=S_{id}+S_{2}-S_{vib}$.

%============\\

%\section{Analytical derivation}
\subsection{Mode coupling Theory}
Mode coupling theory (MCT) is a well-known theory for glass forming liquids. This microscopic theory can qualitatively predict the
dynamics of the glass forming liquid, if the structure of the liquid is known. Many experimental and simulation studies 
have proved 
that these predictions made by MCT are true \cite{gotze-jpcm-10A1}.  
The equation of motion for the intermediate scattering function is given by
\begin{equation}
 \ddot{\bf{S}}(k,t)+{\bf{\Gamma}} \dot{\bf{S}}(k,t)+{\bf{\Omega}}_k^2 {\bf{S}}(k,t)+{\bf{\Omega}}_k^2\int {\bf{\mathcal{M}}}(k,t-t')
 \dot{\bf{S}}(k,t')dt'=0,
\label{fkt-eqn}
\end{equation}
where ${\bf{\Omega}}_k^2=\frac{k^2k_BT}{m}{\bf{S(k)}}^{-1}$, ${\bf{S}}(k,t)$ is the matrix of intermediate scattering function
$S_{\alpha \beta}(k,t)$ and memory function ${\bf{\mathcal{M}}}(k,t)$ can be written as  :
\begin{eqnarray}
\label{memory-fkt}
 [{\bf{\Omega}}^2_k{\bf{\mathcal{M}}}(k,t)]_{\alpha \beta}&=&\frac{1}{2\rho \sqrt{x_\alpha x_\beta}}\sum_{ll'mm'}\int \frac{d\bf{q}}{(2\pi)^3}V_{\alpha lm}({\bf{q,k-q}})\nonumber\\
  &\times&V_{\beta l'm'}({\bf{q,k-q}})S_{mm'}({\bf{|k-q|}})\nonumber\\
 &\times&S_{ll'}(q)\phi_{mm'}(\mid{\bf{k-q}}\mid,t)\phi_{ll'}(q,t),
\end{eqnarray}
where $\phi_{\alpha\beta}(k,t)=\frac{S_{\alpha\beta}(k,t)}{S_{\alpha\beta}(k,0)}$, ${\bf{k-q}}=\bf{p}$ and 
$ V_{\alpha lm}({\bf{q}},{\bf{p}})=[\hat{\bf{k}}.{\bf{q}}\delta_{\alpha m} C_{\alpha l}(q) + \hat{\bf{k}}.{\bf{p}}\delta_{\alpha l}
C_{\alpha m}(p)]$. ${\bf{C}}(k)$ is defined as ${\bf{S}}(k)^{-1}={\bf{1}}-{\bf{C}}(k)$. The static structure factors,
 $S_{\alpha \beta}(k)$ are calculated from
simulation and are defined as,
\begin{equation}
 S_{\alpha \beta}(k)=\frac{1}{\sqrt { N_{\alpha} N_{\beta}}} \sum_{i=1}^{N_{\alpha}} \sum_{j=1}^{N_{\beta}} \exp(-i{\bf{k}}.({\bf{r}}_i^{\alpha}-{\bf{r}}_j^{\beta})).
\label{sq-eqn}
\end{equation}
Solving Eq.(\ref{fkt-eqn}) we can obtain the relaxation time from the decay of $\phi_{\alpha\beta}(k,t)$. The 
temperature dependence of the relaxation time provides us the information of  
 the transition temperature $T_c^{micro}$.

%------------------------------------------------\\
\section{Results and Discussions}
%\subsection{Role of RMPE in dynamics}
As mentioned in the Introduction, many properties of the GCM are different from the conventional glass forming liquids like KA 
model \cite{ikeda2011slow,ikeda_Japan,meanfieldGCM} and they appear more mean-field like.
Here we present a comparative study between the GC and KA models like it has been done before.
However, our analysis primarily focuses on the thermodynamic
properties like the excess entropy and the configurational entropy and their components like pair and higher order terms. We also 
study the correlation between entropy and dynamics.

\begin{figure*}[ht!]
\centering
\subfigure{
 \includegraphics[width=0.40\textwidth]{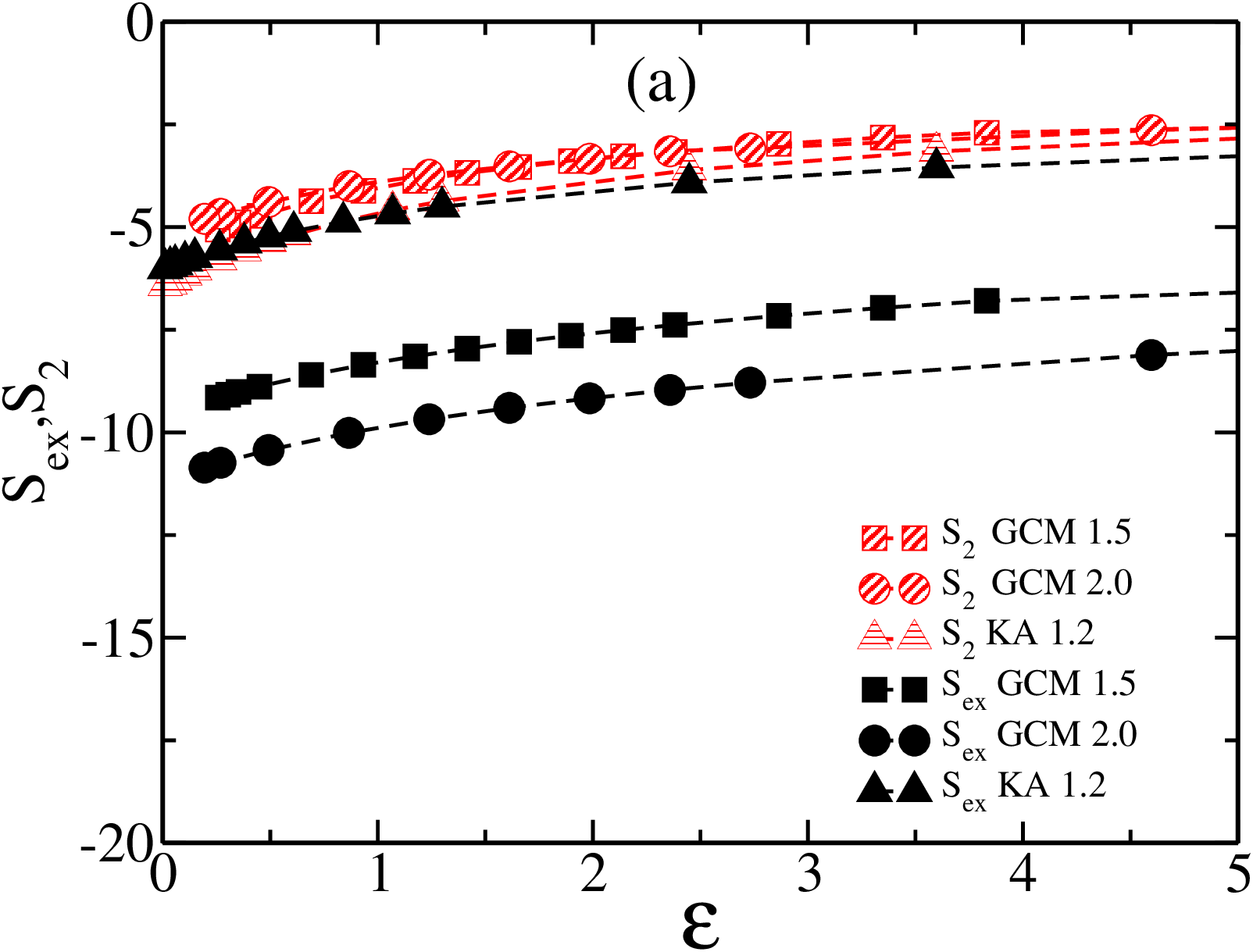}}
 \subfigure{
 \includegraphics[width=0.38\textwidth]{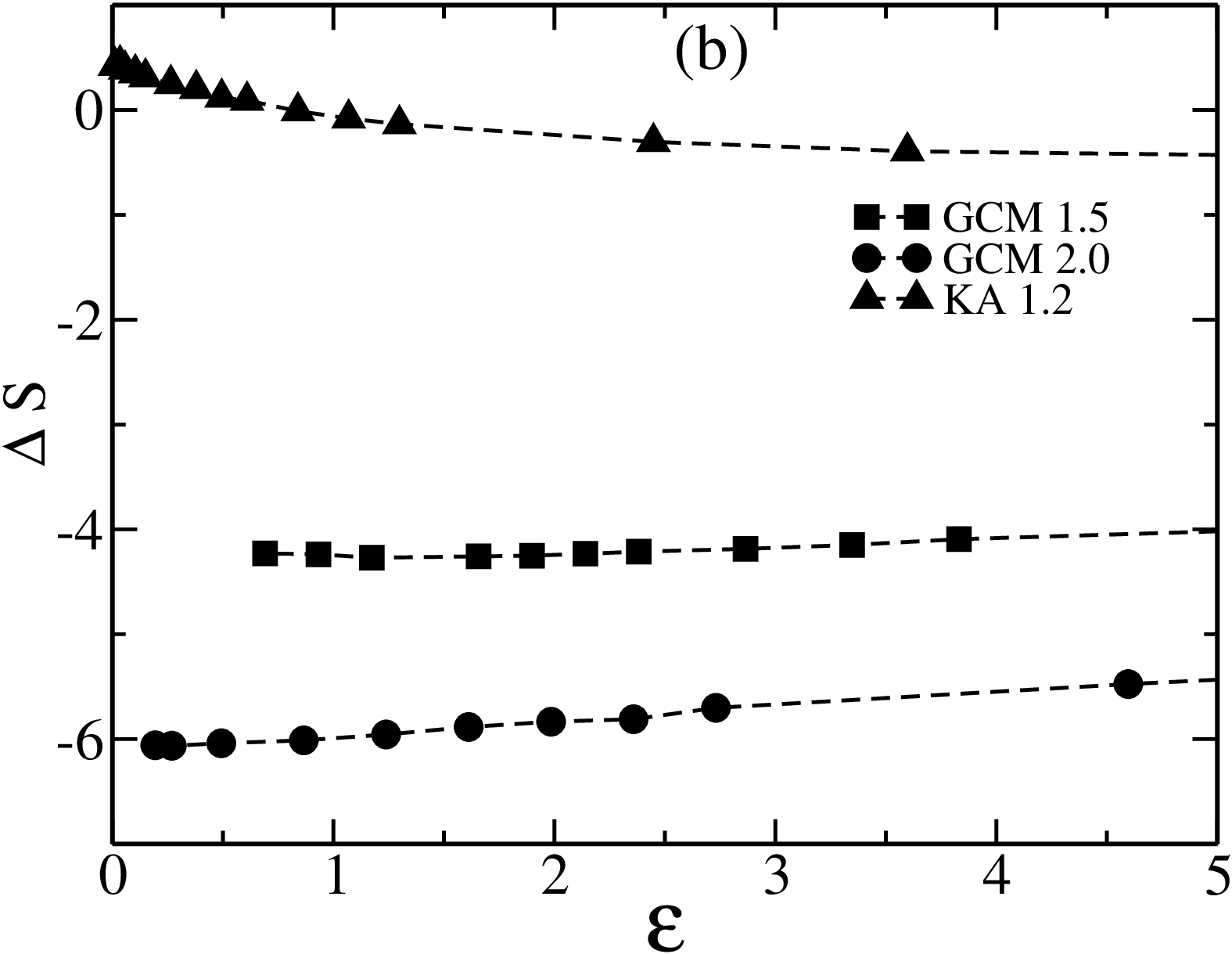}}
\caption{(a) $S_{ex}$ and $S_2$ are plotted as a function of temperature for GCM at $\rho=1.5$ and 2.0 and KA model at $\rho=1.2$.
As the temperature range for 
GCM and KA model are very different, to make a meaningful comparison we plot these entropies against 
$\varepsilon=(\frac{T}{T_c}-1)$.
In KA model there is a crossing between $S_{ex}$ and $S_2$.
(b) $\Delta S$ against temperature for the same models. Only for KA model
$\Delta S$ has positive contribution.}
\label{s2-sex}
\end{figure*}
\subsection{Excess entropy}
First, we study the excess entropy and its different components, the pair, $S_2$ and the higher order terms, $\Delta S$ (Sec. IIIB1, Eq.
\ref{deltas_sexc}).
Our study reveals that unlike in KA model where there is a clear separation of
major contribution to high temperature MCT like dynamics and low temperature activated dynamics from the pair and higher order terms
of the entropy respectively \cite{atreyee_prl,prl_long,onset,unravel}, 
 in GCM that separation does not exist. We plot the $S_{ex}$ and $S_2$ for 
 both KA and GC models as a function of temperature and observe certain stark differences between them.
  Since the temperature range for GCM and KA model are very different, to make
 a meaningful comparison, in the x-axis, we plot $\varepsilon=(\frac{T}{T_c}-1)$(Fig. \ref{s2-sex}).
 Unlike in KA and other simple glass
forming liquids where at high temperature $S_2$ 
contributes to $80\%$ of $S_{ex}$, in GCM we find that even at temperature $T\simeq 10\times T_c$
the dynamics is dominated by $\Delta S$ and 
contribution of $S_2$ is only $34\%$ for $\rho=1.5$  and $29\%$
for $\rho=2.0$ (Table-\ref{s2_contribution}). Thus 
it implies that in GCM many-body correlations
dominate the dynamics even at high temperatures and this contribution increases with density. 
 Note that earlier studies have shown that the unstable modes which are characteristics of high temperature dynamics and disappear at
$T_c$ are delocalized  for GCM whereas they are localized for KA model \cite{meanfieldGCM}. This can be explained from our 
entropy calculation. At high temperatures dominance of many-body correlation in GCM implies delocalized mode whereas 
dominance of pair correlation
in KA model implies localized mode. Large absolute value of $\Delta S$ in GCM also suggests strong cooperative motion and can be
connected to the earlier findings of larger value of $\chi_4$ when compared to KA model \cite{meanfieldGCM}.
%=======================Table S2 Sex==============================================================================================
\begin{table*}[ht!]
\caption{ \it{$S_{ex}$, $S_2$ and $\Delta S$ at high temperature where $T\sim 10 \times T_c$. In
KA model $S_2$ contributes to $80\%$ of $S_{ex}$, whereas in GCM at $\rho=1.5$ it is $34\%$ and at $\rho=2.0$ the contribution is $29\%$.}}
 \centering
\begin{tabular}{ l | r | r | r | r | r | r  }
 \hline
   & $T_c$ & $T\sim 10\times T_c$ & $S_{ex}$ & $S_2$ & $\Delta S$ & $\frac{S_2}{S_{ex}}\%$\\
\hline 
KA               &0.435                & 5.00 & -2.62092 & -2.14422 & -0.4767 & 82      \\
GCM ($\rho=1.5$) &2.07$\times 10^{-5}$ &2.00$\times 10^{-4}$& -5.73985& -1.96565 & -3.7742 & 34  \\

GCM ($\rho=2.0$) &2.68$\times 10^{-6}$ &3.00$\times 10^{-5}$& -7.04557& -2.05902 & -4.9866 & 29  \\
\hline

\end{tabular}
\label{s2_contribution}
\end{table*}

%=================================================================================================================================

In KA model with decrease in temperature the $S_2$ and $S_{ex}$ undergo a crossing and the many-body contribution to the entropy 
becomes positive. Recently we have shown that this crossing marks the onset temperature \cite{onset} and the positive value of
$\Delta S$ is associated with the activated dynamics as it leads to the increase in entropy and thus speed up of dynamics 
\cite{onset,atreyee_prl, unravel}. 
Here we find that in GCM, $S_2$ and $S_{ex}$ never undergo a crossing and 
$\Delta S$ is never positive (Fig. \ref{s2-sex}). Thus we cannot predict an onset temperature from the entropy.
It has been earlier observed that for hard sphere system in higher dimensions
$\Delta S$ remains negative for a wider density regime \cite{truskett_2008}. In 3D, $\Delta S=0$ (marking a transition from negative 
to positive value)
at the freezing density whereas in 4D the density where $\Delta S=0$ is higher than the freezing density and in 5D the $\Delta S$
remains negative much above the freezing density and the absolute value is much higher than that obtained for 3D and 4D systems. 
Similar difference between freezing point and $\Delta S=0$ point has also been observed in GCM \cite{rmpe_saija2}.
These studies reveal that as we 
go to mean-field like systems the negative value of  $\Delta S$ persists for a wider range of density or temperature and the absolute
 value of $\Delta S$ increases. Hence the fact that $\Delta S$ has a large negative value in GCM supports the earlier findings that
 GCM exhibits mean-field like behaviour \cite{meanfieldGCM} usually observed in higher dimensional systems.

As mentioned before in our earlier study we have connected
the positive value of $\Delta S$ to the activated dynamics \cite{atreyee_prl,prl_long,unravel}. 
The positive value of $\Delta S$ implies higher order correlations increase the entropy, similarly activated dynamics
which is many-body in nature is supposed to allow the system to explore more configurational space.
Thus a negative value of $\Delta S$ in GCM predicts suppression of activation.
This suppression of activation has already been reported by Coslovich {\it et al.} % Ikeda and Miyazaki 
where they have
shown that in GCM the van Hove correlation function ($G_s(r,t)$) does not show any bimodal distribution even at very low 
temperatures, which implies no hopping like motion \cite{meanfieldGCM}. They have claimed that in the landscape picture
this is due to the higher value
of energy barriers \cite{meanfieldGCM}. 

To summaries, $\Delta S$ in KA model has small values and undergoes a sign change whereas absolute value of $\Delta S$ in GCM 
is large and remains negative.
Thus many-body correlation in KA model is 
weak and it contributes primarily at low temperatures
to the activated dynamics. In contrast, many-body correlation is strong in GCM, has contribution both at high and low temperature
regimes but it primarily contributes to slowing down of the dynamics.
It is possible
that there can be some parts of the many-body contribution which speeds up the dynamics and some parts which slows it down as 
reported by Coslovich {\it et al.} \cite{meanfieldGCM} where they have found the presence of both cooperative and incoherent 
many-body modes \cite{meanfieldGCM}. However in the present study such 
separation in the calculation of $\Delta S$ is not possible.

\begin{figure*}[ht!]
\centering
\subfigure{
 \includegraphics[width=0.38\textwidth]{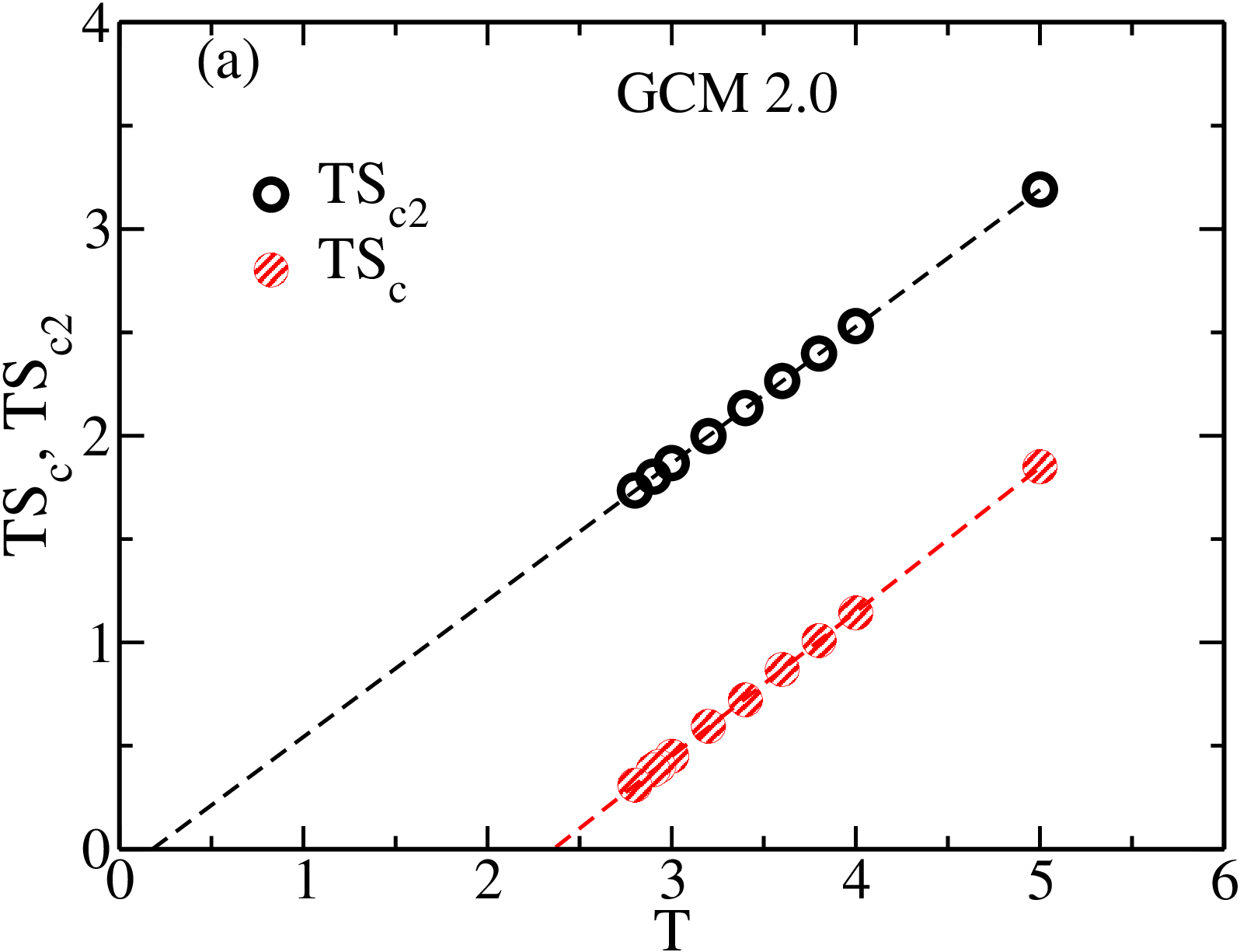}}
 \subfigure{
 \includegraphics[width=0.38\textwidth]{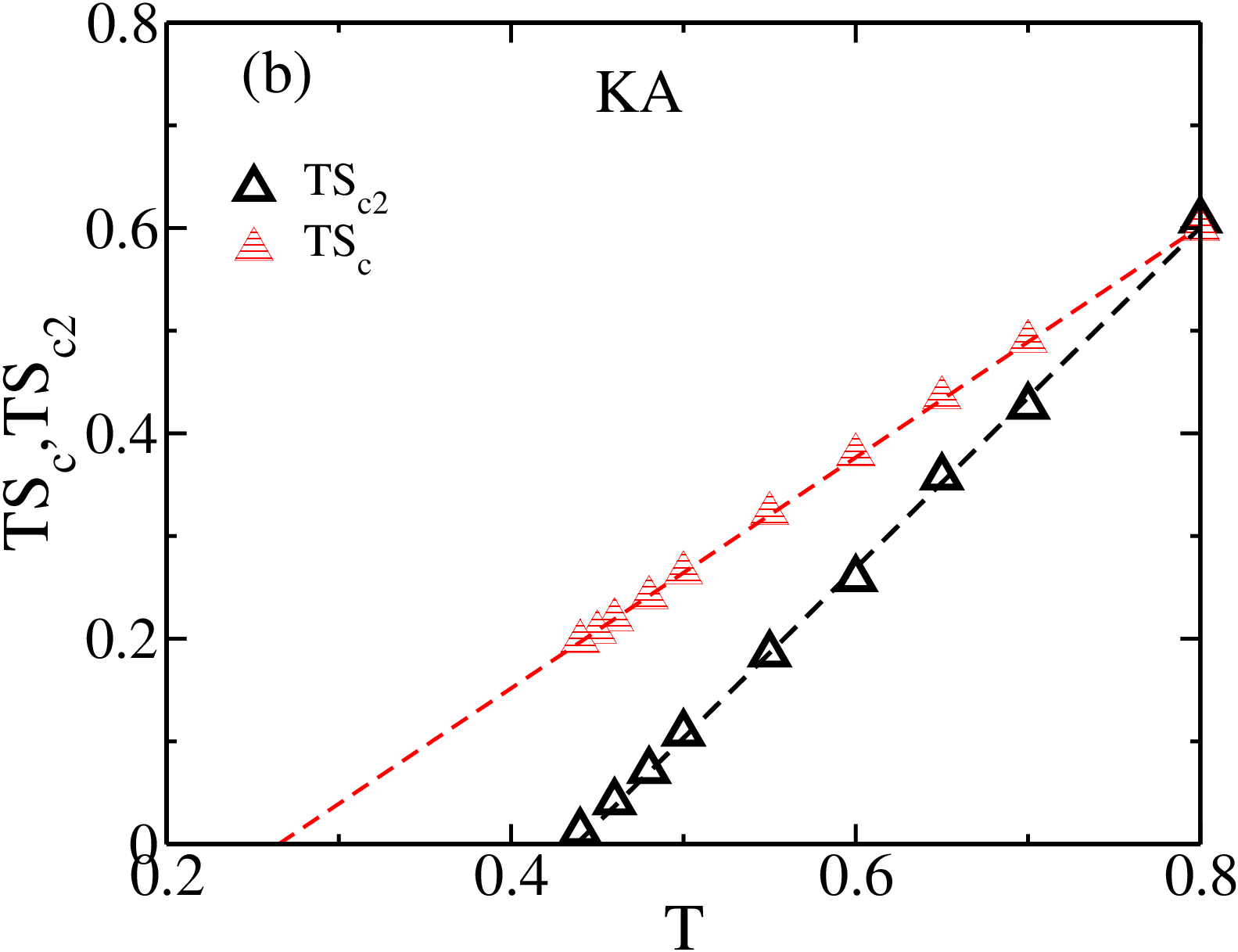}}
\caption{(a) Temperature dependence of $TS_{c}$ and $TS_{c2}$ for GCM at $\rho=2.0$. Both show linear behaviour.
The Kauzmann temperature, $T_K$ and
pair Kauzmann temperature, $T_{K2}$ obtained from the extrapolation of the lines. To plot $TS_{c}$ and $TS_{c2}$ in the same figure,
we have divided $TS_{c2}$ by 10. (b) Same as in (a) but for the KA model. In GCM the two lines run almost parallel and $T_{K2} < T_K$,
whereas in KA model the slopes are different, the lines cross each other and $T_{K2} > T_K$.}
%The values are given in Table-\ref{Tc_value}.}
\label{Tk_Tk2_value}
\end{figure*}

\subsection{Configurational entropy}
Next we study the different contributions to the configurational entropy, $S_c$. Since this calculation is time consuming, for 
GCM we concentrate only at $\rho=2.0$.
Note that for glass forming systems $TS_c$ vs T is found to be linear \cite{srikanth,atreyee_prl,prl_long} and the extrapolation of
the linear fit gives us a measure of the Kauzmann temperature, $T_K$ where the extrapolated $S_c$ vanishes. In our earlier work we have shown
that $TS_{c2}$ vs T also shows a linear behaviour and predicts a transition temperature, $T_{K2}$ where the $S_{c2}$ vanishes 
\cite{unravel,role_pair}. In Fig. \ref{Tk_Tk2_value}a and \ref{Tk_Tk2_value}b we plot the $TS_c$ and $TS_{c2}$ against temperature
 both for GC and KA models 
respectively. We find that similar to KA model the $TS_c$ and $TS_{c2}$ vs T plots in GCM are linear.
However, the main difference is that in GCM the two lines run almost parallel and $T_{K2} < T_K$, whereas in KA model the slopes 
are different, the lines cross each other and $T_{K2}>T_K$.  This is related to the observation mentioned before (Fig. \ref{s2-sex}a)
that in GCM, $S_2$ and $S_{ex}$ do not cross but in KA model they cross at the onset temperature.
For conventional glass forming liquids like KA model \cite{atreyee_prl,unravel,role_pair}
we have also reported that $T_{K2}\simeq T_c$.  In case of GCM $T_{K2}<< T_c$.
The possible explanation for this is that $T_c$ marks the disappearance of 
high temperature dynamics. For systems where $S_2$ provides a dominant contribution at high temperatures the corresponding configurational 
entropy vanishes at $T_c$.
 This does not appear to be the case 
in GCM. As discussed before the high temperature
non-activated dynamics in this system is dominated not by the pair but by the many-body correlations. 
 Thus the disappearance
of the high temperature dynamics around $T_c$ has no connection with the vanishing of the $S_{c2}$ and hence
$T_{K2} \neq T_c$.

\subsection{Adam-Gibbs relation}
\begin{figure*}[ht!]
\centering
\subfigure{
\includegraphics[width=0.45\textwidth]{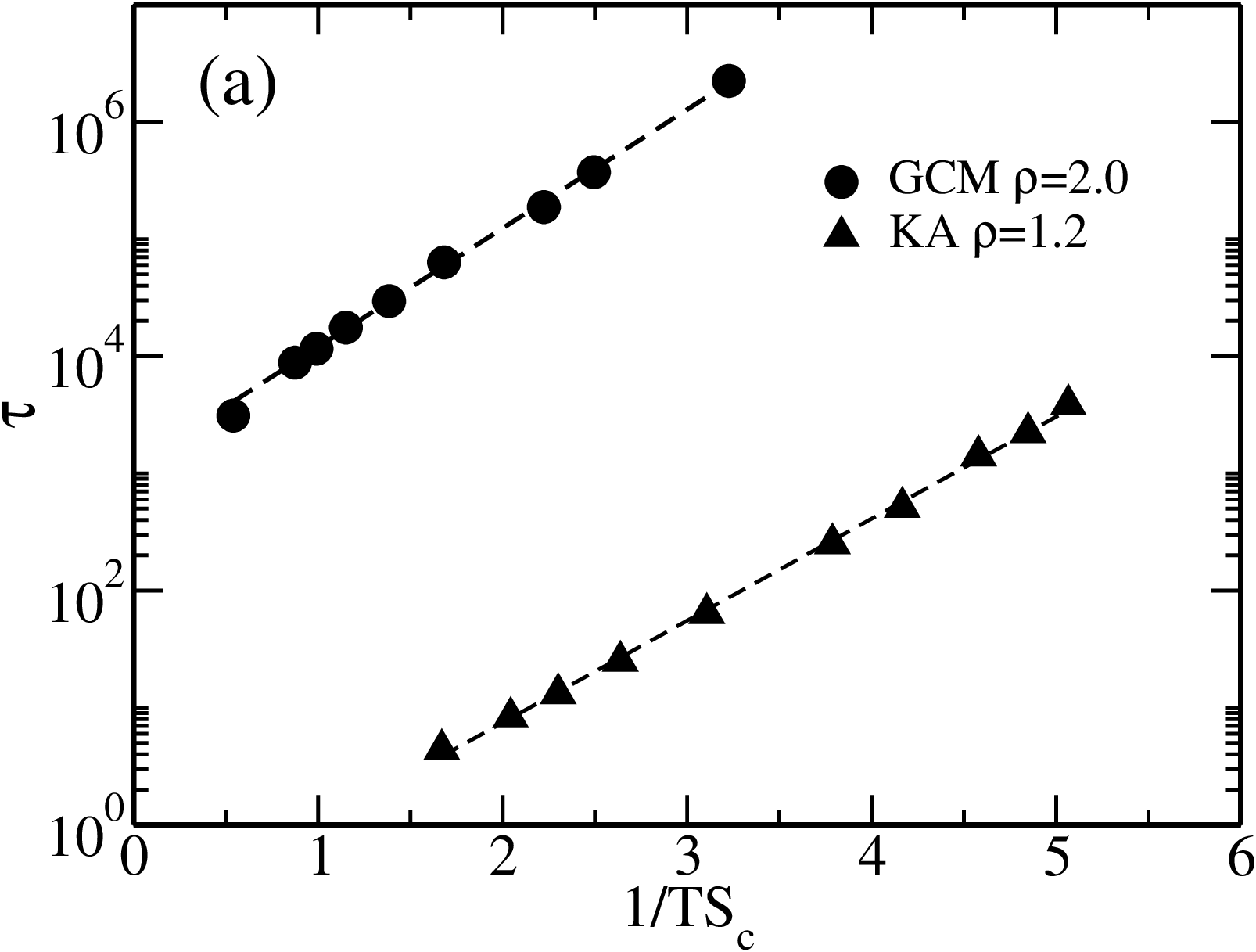}}
\subfigure{
\includegraphics[width=0.45\textwidth]{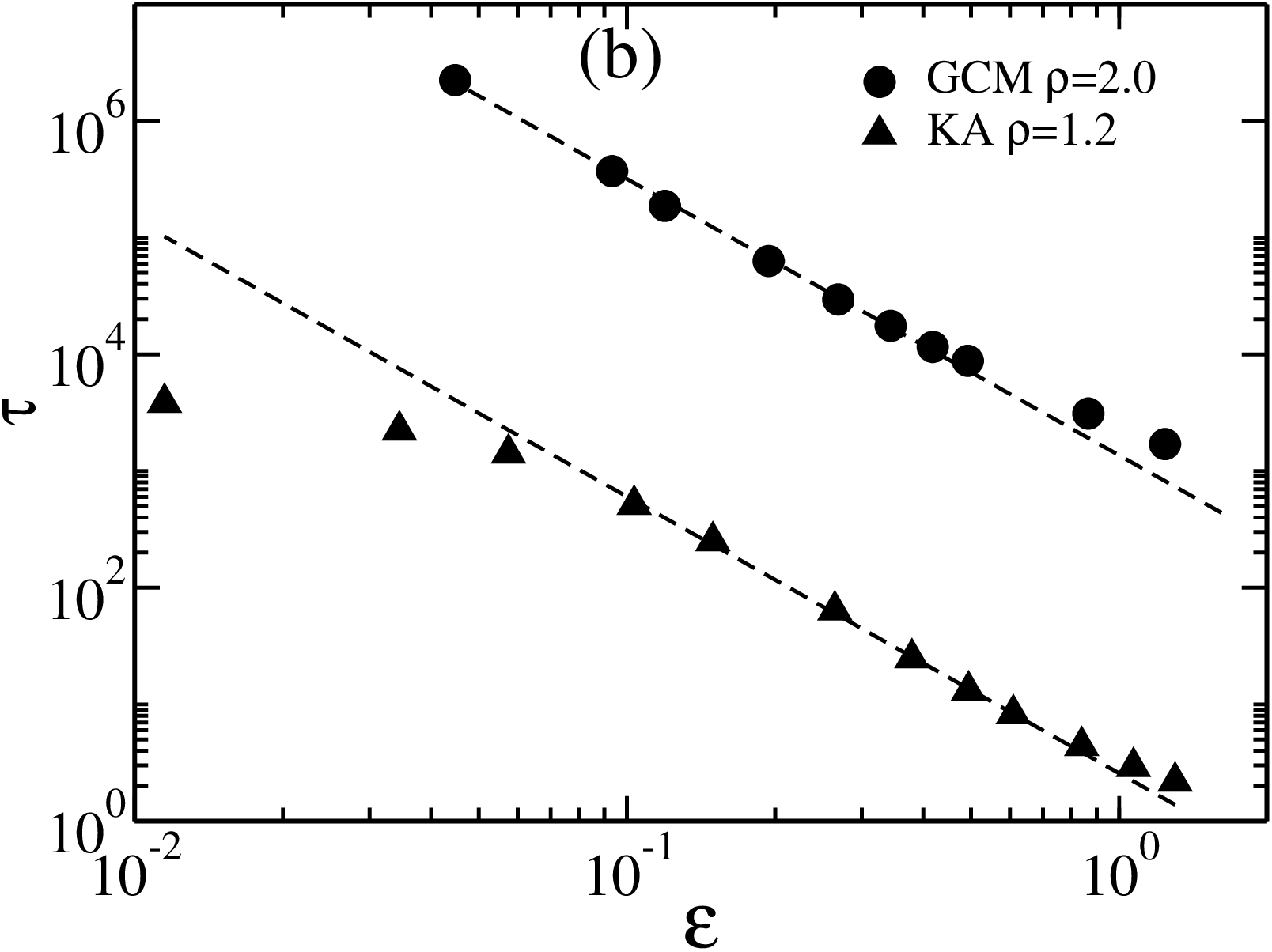}}
\subfigure{
\includegraphics[width=0.45\textwidth]{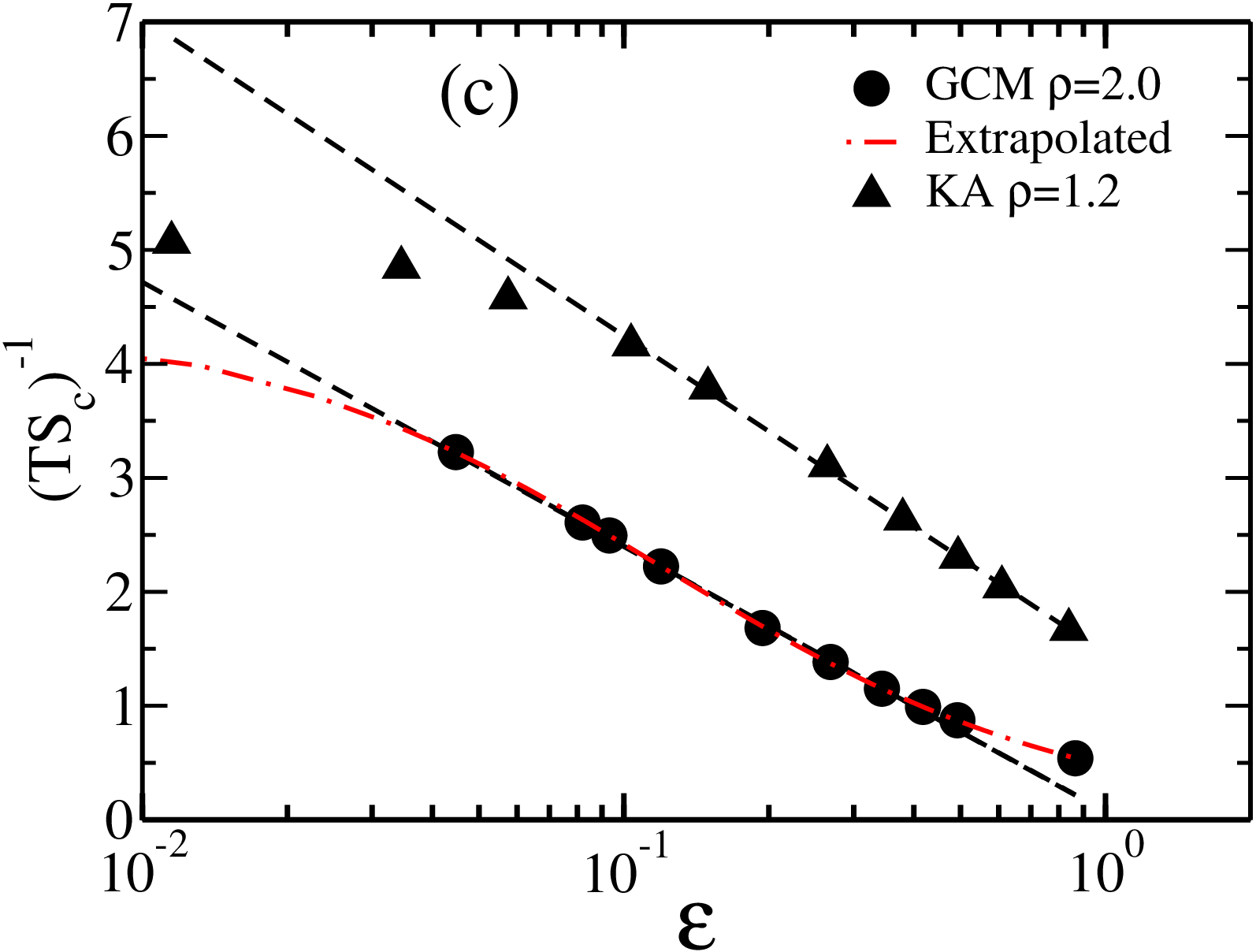}}
 \caption{Comparative study of GCM at $\rho=2.0$ and KA model at $\rho=1.2$. (a) Adam Gibbs plot for the relaxation time 
 ($\tau$ vs $\frac{1}{TS_c}$). Both follow AG relation. (b) The MCT power-law 
 behaviour of relaxation time, $\tau$ vs $\varepsilon$ where $\varepsilon=(\frac{T}{T_c}-1)$. The dashed lines are MCT fit 
 ($\tau\propto (T-T_c)^{-\gamma}$).
 (c) $(TS_c)^{-1}$ against $\varepsilon$. The dashed lines are MCT like power-law fit (Eq. (\ref{ag_mct_overlap})). The dashed-dot 
 line is the 
 extrapolated value of $(TS_c)^{-1}$ obtained from Fig. \ref{Tk_Tk2_value}a.
 For better comparison in GCM model $(TS_c)^{-1}$ is multiplied 
 by $10^{-6}$. We find that in (b) the relaxation time, $\tau$ and in (c) $(TS_c)^{-1}$ 
  follow power-law behaviour in the same temperature regime. However for GCM when compared to KA model 
 this regime is shifted to lower temperatures.
For GCM the simulated values of $\tau$ and $(TS_c)^{-1}$ do not show any departure from power law.
 The extrapolated $(TS_c)^{-1}$ (shown in (c)) can predict a departure from power law and a
 transition to activation dominated dynamics.}
 \label{tau_mct_powerlaw}
\end{figure*}

As discussed in the Introduction it is believed that the low temperature dynamics for glass forming liquids is activated in nature and 
the relaxation time, $\tau$ is related to the configurational entropy, $S_c$ via AG relation:
\begin{equation}
\tau(T)=\tau_{o}\exp\left(\frac{A}{TS_{c}}\right),
\label{ag}
\end{equation}
where A is the AG coefficient.
Since the GCM shows a suppression of activation \cite{ikeda2011slow,meanfieldGCM} we expect the AG relation to be 
violated in this system. To our surprise
we find that in GCM the AG prediction of connection between dynamics and entropy holds as shown in Fig. \ref{tau_mct_powerlaw}a.
This is the first time it is shown that for systems where activated dynamics is clearly suppressed the AG relation holds.
In our
earlier study we have already shown that AG relation holds not only at low temperatures where activated dynamics is dominant but 
also at reasonably high temperature, where the dynamics is still described by MCT \cite{unravel}. Thus we claimed 
that there 
exists a non-activated contribution to AG and our present finding supports this 
argument. 
Note that in an earlier study it has been shown that in the 4D system AG relation holds but there has been no discussion about the 
suppression of activated dynamics \cite{shila_4d}. 
\\%For KA model we have also shown that there is an overlap between the AG and the ....\\
  
%=======\\
%============\\

\subsection{AG and MCT overlap regime}
In order to further understand this non-activated contribution to AG, we analyze the MCT power law behaviour of the relaxation time
and the configurational entropy.
In Fig.\ref{tau_mct_powerlaw}b we plot the relaxation time for GCM and KA models. As reported earlier, like KA model the relaxation
time, $\tau$ of GCM follows MCT like power law behaviour
and predicts a transition temperature $T_c$ (Table-\ref{Tc_value}). 
For most of the glass forming liquids like KA model the range of this regime is $0.1 < (\frac{T}{T_c}-1)<1.0$ \cite{unravel}. 
In contrast the power law regime in GCM
 is shifted towards lower temperatures. For GCM 
we do  not find any deviation from MCT power law till the temperature we have studied ($(\frac{T}{T_c}-1)\simeq 0.0448$). 
Thus from this figure we cannot comment if the MCT divergence is real or avoided.
 Note that according to microscopic MCT calculation 
this  power law regime appears at a lower temperature \cite{szamel-pre}. It will be interesting to understand if this shift is also a 
mean-field effect. But this is beyond the scope of the present study.

Although fitting relaxation time and diffusion coefficient to MCT power law behaviour is done routinely, the study of the power law
behaviour of the configurational entropy is not a standard protocol. In a recent work on KA model we have shown that there is an
overlap between AG and MCT regime \cite{unravel}. In this common regime we can write,
\begin{equation}
 \frac{A}{TS_c}\propto \ln(\frac{T}{T_c}-1).
 \label{ag_mct_overlap}
\end{equation}
Thus the study of the power law behaviour of $(TS_c)^{-1}$ is the best way to understand this regime.
We find that similar to KA model in GCM there is an overlap between AG and MCT regime.  
%In Fig. \ref{tau_mct_powerlaw}c we plot $(TS_c)^{-1}$ vs. $\ln(\frac{T}{T_c}-1)$ for GCM and for comparison we also plot the same for KA 
%model.
Like in KA model
the $(TS_c)^{-1}$ vs $\ln (\frac{T}{T_c}-1)$ in GCM follows a linear behaviour (Fig. \ref{tau_mct_powerlaw}c). 
However unlike KA model and similar to what we observe
for the relaxation time, this region is shifted to a lower temperature (Fig. \ref{tau_mct_powerlaw}b).

%=================\\
%Note that both for KA model and GCM, 
A consequence of the validity of both AG and MCT relation is that both 
$TS_c$ vs. T  and  $(TS_c)^{-1}$ vs $\ln (\frac{T}{T_c}-1)$
 show a linear behaviour.
The former linear behaviour predicts a vanishing of $S_c$ at $T_K$ and the latter predicts that it vanishes at $T_c$. In KA model
we have shown that these two contradicting behaviour appears because a part of $S_c$ vanishes at $T_c$ and is responsible for the
predicted divergence like behaviour at $T_c$. Interestingly this part of the entropy, namely the pair part
is connected to the high temperature dynamics.
The other part of $S_c$, {\it ie.} $\Delta S$ survives and provides a finite value to $S_c$
below $T_c$. This makes the divergence at $T_c$ an avoided one and leads to the departure from linearity of the $(TS_c)^{-1}$ vs
$\ln (\frac{T}{T_c}-1)$ plot (Fig. \ref{tau_mct_powerlaw}c).
 This departure also 
marks a transition to an activation dominated regime, although the onset of activation happens at a higher temperature,
$T_{onset}$ \cite{onset}.
Note that similar to KA model in GCM we find both $TS_c$ vs T and $(TS_c)^{-1}$ vs $\ln(\frac{T}{T_c}-1)$ to be linear, thus 
predicting two vanishing temperatures for $S_c$, one at $T_K$ and the other at $T_c$, respectively.
 Till the lowest temperature studied here we do not find any departure from linearity of the 
$(TS_c)^{-1}$ vs $\ln (\frac{T}{T_c}-1)$ plot.
 Thus we can not comment
if the transition at $T_c$ is real or avoided. However if we plot the extrapolated value of $TS_c$ as obtained 
from Fig. \ref{tau_mct_powerlaw}a, we find that $(TS_c)^{-1}$ vs $\ln (\frac{T}{T_c}-1)$ shows a departure from 
linearity (Fig. \ref{tau_mct_powerlaw}c). If we trust the extrapolation if not till $T_K$ but at least till some temperature 
which is lower than that studied here, then this 
departure implies two things. First, not the whole $S_c$ but
some components of $S_c$, most probably the ones related to the high temperature dynamics 
vanishes at $T_c$ and second, at some temperature above $T_c$, the remaining part of $S_c$ becomes 
positive suggesting the presence of activation and 
around $T_c$ the activated dynamics becomes dominant.   
Note that unlike in KA model in GCM it is not the pair part of $S_c$ that vanishes at $T_c$. 
As mentioned before, this is because $S_2$ is not the dominant contributor to high temperature dynamics. 
From the present study we cannot specify till what order in $S_{ex}$ contributes to high temperature dynamics. However the departure
from linearity seen in Fig. \ref{tau_mct_powerlaw}c and thus
the prediction of a transition to 
activated dynamics around $T_c$ is similar to the earlier observations \cite{atreyee_prl,unravel,role_pair}.

\subsection{Activation dominated regime}
Next we show that
our study further reveals that the activation
dominated regime in GCM is very small. In Table-\ref{Tc_value} we %plot
have given the $T_c$, $T_K$ and also $T_{VFT}$ values, where %$T_c$ is 
%obtained from power law fit, $T_K$ is obtained from extrapolation of $TS_c$ and 
$T_{VFT}$ is obtained from fitting $\tau$ to 
$\tau \sim \tau_0 exp\left[ \frac{1}{K_{VFT}(T/T_{VFT}-1)}\right] $ form. We have also tabulated the 
$\frac{T_c-T_K}{T_K}\%$ and 
$\frac{T_c-T_{VFT}}{T_{VFT}}\%$ values 
for KA and GCM systems. First we find that $T_K$ and $T_{VFT}$ are very close which is a reflection of the validity of the AG 
relation \cite{srikanth}. We also find that the difference between $T_c$ and $T_K$/ $T_{VFT}$ is much smaller in GCM 
than in KA model (Table-\ref{Tc_value}).
This implies that the activation dominated regime in GCM is much smaller than that in KA model.
Although our study predicts a transition to activated dynamics and that
the regime of activated dynamics to be small it can not predict the degree of contribution of the activation to the total dynamics.\\

%===========\\
\begin{figure}[ht!]
\centering
 \includegraphics[width=0.38\textwidth]{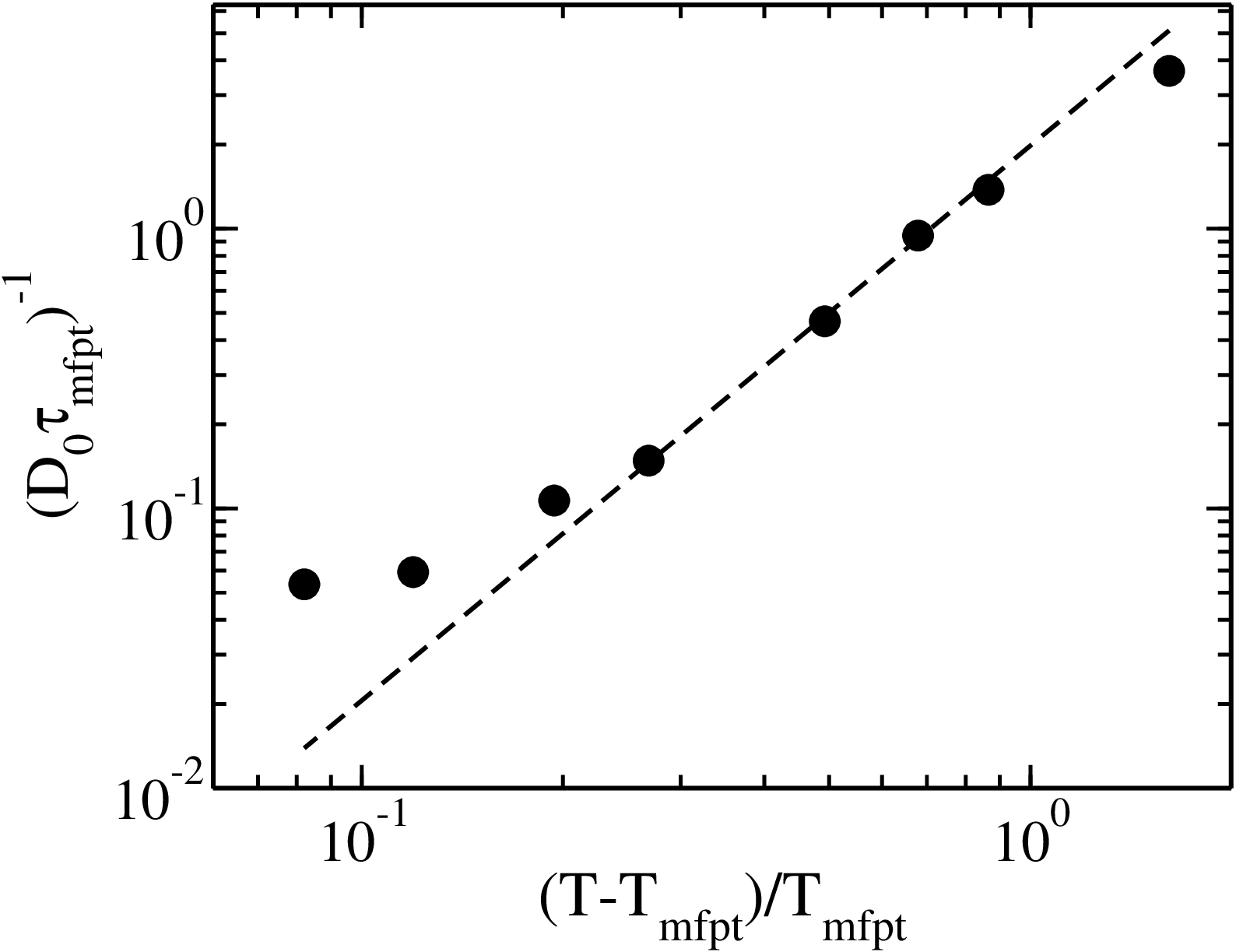}
\caption{The power law dependence of $(D_0\tau_{mfpt})^{-1}$ predicts a transition temperature $T_{mfpt}$, which is close to the 
$T_c$ in scaled unit. The dashed line is the power law fit.}
\label{mfpt_t_tc}
\end{figure}

%==================10 oct=====================================

\begin{table*}[ht!]
\caption{ \it{The values for different dynamic and thermodynamic transition temperatures for
KA at $\rho=1.2$ and GCM at $\rho=2.0$. $T_K\simeq T_{VFT}$ implies AG relation is valid. $\frac{T_c-T_K}{T_K}\%$ and 
$\frac{T_c-T_{VFT}}{T_{VFT}}\%$ are less for GCM suggesting narrow activation dominated regime.}}
%The deviation between the two temperatures are also given \cite{role_pair,kuni-jphysJapan}. }}
 \centering
\begin{tabular}{ l | r | r  }
 \hline
   &$KA (LJ \rho=1.2)$ &  $GCM (\rho=2.0)$\\
%   &$KALJ (\rho=1.2)$ &NTW ($\rho=1.655$) & $GCM (\rho=1.5)$ & $GCM (\rho=2.0)$\\
\hline 
$T_c$ &0.435 & 2.68$\times 10^{-6}$\\
%$T_{MCT}^{theory}$ &0.887 & 3.17$\times 10^{-6}$\\
$T_K$         &0.27 &  2.36 $\times 10^{-6}$\\
$T_{VFT}$ & 0.28 & 2.31$\times 10^{-6}$\\
$T_{mfpt}$ & 0.428$\pm$ 0.022 & 2.66$\pm$ 0.01\\
\hline
%$\frac{T_{MCT}^{theory}-T_{MCT}^{sim}}{T_{MCT}^{sim}}\%$ &104$\%$ &18$\%$\\
%\hline
$\frac{T_c-T_K}{T_K}\%$ & 61.11$\%$& 13.56$\%$\\
\hline
$\frac{T_c-T_{VFT}}{T_{VFT}}\%$ & 55.36$\%$& 16.02$\%$\\
\hline

\end{tabular}
\label{Tc_value}
\end{table*}
%===========================Mean field calculation==================================================================
\subsection{Mean-field theory (MFT) approach}

Recently we have developed a mean-field like theory which can describe the dynamics of a collection of interacting
particles in terms of a collection of non-interacting particles in an effective potential \cite{role_pair}. 
The effect of the interaction between the 
particles are absorbed in this effective potential at a mean-field level.
Below we provide a sketch of the derivation with few important equations. The details are given in Ref. \cite{role_pair}.
Starting from the
Fokker-Planck equation we derived the Smoluchowski equation with an effective potential $\Phi(r)$. Using the dynamic density functional
approach \cite{ramakrishnan-yossuf} we obtained the effective caging potential as,
\begin{equation}
 \Phi(r)=-\frac{1}{2}\int \frac{d{\bf q}}{(2\pi)^3}\rho C^2(q)S(q)e^{-q^2r^2/3}.
\label{mean-field_potential}
\end{equation}
Here $C(q)$ is the direct correlation function and $S(q)$ is the static structure factor of the liquid.
Note that the caging potential depends only on the equilibrium pair correlation function. Next we calculated the mean first passage 
time, the time required to escape from the effective potential which leads to caging of the particles,
 \begin{equation}
  \tau_{mfpt}=\frac{1}{D_0}\int_0^{r_{max}} e^{\beta \Phi(y)} dy\int_0^y e^{-\beta \Phi(z)} dz,
  \label{mfpt}
 \end{equation}
where $D_0=k_BT/\zeta$ and  $\zeta$ is the coefficient of the friction of the system and $r_{max}$ is the range of localization 
potential $\Phi(r)$.
 As done earlier for other glass forming liquids \cite{role_pair},  we calculate the mean first passage time
 in GCM. Note that the range of temperatures in GCM is much smaller compared to standard glass forming systems. This leads to 
 numerical problems in the calculation of $\tau_{mfpt}$ as the temperature is in the exponential. However we can scale the potential
 and temperature in such a way that the temperature range moves to higher values. For this part of the calculation, we run simulations
 where $\epsilon =10^6\epsilon_0$, the temperature and the time are scaled as $T=T\times 10^6$ and $\Delta t=\Delta t\times 10^{-3}$.
 Although the time scale changes,  
the static properties like radial distribution function, $g(r)$ and structure factor, $S(q)$ remain same as in the original system and 
dynamics gets appropriately scaled. 
In Fig. \ref{mfpt_t_tc}, we show that $\tau_{mfpt}$ follows a power law behaviour and the transition temperature $T_{mfpt}\simeq T_c$ 
in the scaled unit (Table-\ref{Tc_value}). Note that for KA model and other systems $T_{mfpt}\simeq T_{K2}\simeq T_c$.
For GCM as discussed
before $T_{K2}$ is much smaller as $S_2$ is not the dominant contributor to the high temperature dynamics. However, we find that 
although $S_{2}$ cannot predict the full high temperature dynamics the present mean-field theory using the same pair correlation
as used for the calculation of $S_{c2}$ can.

%\clearpage

\section{conclusion}
In this work we present a comparative study between the GC and KA models.
The work is similar in spirit to that presented earlier by other groups where it was
found that the dynamic properties in GCM are quite different from that in KA model and are more mean-field 
like \cite{ikeda2011slow,ikeda_Japan,meanfieldGCM}.
 However, in this work, we focus on the calculation of the entropy and 
its components and the study of the correlation between entropy and dynamics. 
Our study supports the conclusions made in the earlier studies and also makes some new predictions.

The excess entropy which is the loss of entropy of the liquid due to
its correlations can be broken up into pair and higher order terms \cite{green_jcp,raveche,Wallace}. 
For standard glass former like KA model at high temperatures the 
pair part of the excess entropy ($S_2$) contributes to 80$\%$ of the total excess entropy ($S_{ex}$) \cite{atreyee_prl,BORZSAK1992227}.
 Thus high temperature dynamics is dominated by 
two-body correlation. At high temperatures $S_2$ is larger than $S_{ex}$ but with decrease in temperature they undergo a crossing 
which marks the onset temperature \cite{onset}.
The RMPE ($\Delta S$) undergoes a sign change and also a role reversal \cite{atreyee_prl}.
For KA model we have observed that small negative values of $\Delta S$ at high temperatures has very little contribution to the dynamics
whereas small positive value of $\Delta S$ has a large contribution to the low temperature dynamics and has been connected
to activation \cite{atreyee_prl}.
In GCM the scenario is quite different.  
At high temperatures the contribution of $S_2$ to $S_{ex}$ is only $\sim 30\%$.  
 Thus in GCM unlike in KA model the high temperature dynamics is
dominated by many-body correlations. The $S_2$ in GCM is always higher than $S_{ex}$ and they do not undergo any crossing.
Thus we cannot predict an onset temperature from entropy.
The RMPE does not undergo a sign change and no role reversal.
 The absolute value of RMPE in GCM is much larger than that in KA model, thus
predicting larger contribution of many-body correlations which is similar to the observation of high value of $\chi_4(t)$ 
in GCM \cite{meanfieldGCM}. Also
 negative $\Delta S$ value predicts 
suppression of activated motion which has been reported earlier from the study of van Hove correlation function \cite{meanfieldGCM}.

Although there is suppression of activation we find that the AG relation in GCM holds over a wide temperature regime.
 As far as our knowledge this is the first system where both suppression of activation and validity of
AG relation is reported simultaneously. 
In our earlier study on KA model we suggested that observed overlap between AG and MCT regime implies that there is a non-activated 
contribution to AG \cite{unravel}.
 Our present finding strengthens our earlier hypothesis.

 Validity of AG relation implies that $S_c$ vanishes at $T_K$ whereas MCT like power law behaviour of $(TS_c)^{-1}$ suggest
 that $S_c$ vanishes at $T_c$. For KA model we have shown that this apparently contradicting behaviour arises as part of $S_c$  
 vanishes at $T_c$. Note that $T_c$ marks the disappearance of high temperature dynamics and in accordance with that it is the 
 pair part of the $S_c$, $S_{c2}$ that disappears at $T_c$.
 Around but above $T_c$, $\Delta S$ becomes positive 
 which provides a finite value to $S_c$ even when $S_{c2}$ vanishes. This leads to the breakdown of the power law behaviour of
 $(TS_c)^{-1}$ and the transition predicted at $T_c$ is avoided. From our earlier analysis of KA model we can say that 
 in GCM the observed power law behaviour of $(TS_c)^{-1}$
 implies that some part of $S_c$ vanishes at $T_c$. Note that unlike in KA model in GCM $S_{c2}$ does not vanish at $T_c$ as
 $S_2$ is not the dominant contributor to high temperature dynamics. Most likely the configurational entropy summed up till some 
 higher order disappears at $T_c$. From our extrapolated data of entropy we find that at lower
 temperatures there is a breakdown of the MCT power law behaviour of $(TS_c)^{-1}$ suggesting that the remaining part of
 $S_c$ becomes positive and the system makes a transition to activated dynamics.

Using a recently developed MFT \cite{role_pair}, which could predict the MCT transition temperature $T_c$ for standard glass former,
we show that we can predict the $T_c$ in GCM. Note that this model requires only the information of the pair correlation function to 
describe the dynamics.

We would like to conclude by saying that the present study involving primarily the thermodynamical quantities can predict the earlier
observations made from the study of the dynamics \cite{ikeda2011slow,meanfieldGCM} and also makes some new 
predictions.
 Also we will like to mention that 
in GCM instead of breaking up the entropy into pair and higher order terms if we could break the entropy into high temperature and 
low temperature contributions then we believe that the results would have been similar to that obtained in KA model in terms of 
pair and higher order contributions.

%\bibliographystyle{jcp.bst}
%\bibliography{ref_gcm}

\clearpage
\end{document}